\documentclass[aps,prd,twocolumn,preprintnumbers,superscriptaddress,nofootinbib,amsmath,amssymb]{revtex4}

\usepackage{graphicx}
\usepackage{subfigure}
\usepackage{epstopdf}
\usepackage{dcolumn}
\usepackage{amsmath}
\usepackage[normalem]{ulem}
\usepackage[utf8x]{inputenc}

\usepackage{blindtext}
\usepackage{xcolor}

\begin{document}

\title{Dependence of the static quark free energy on $\mu_B$ and the\\
  crossover temperature of $N_f = 2+1$ QCD}

\author{Massimo D'Elia}\email{massimo.delia@unipi.it}
\affiliation{Dipartimento di Fisica dell'Universit\`a di Pisa, Largo B.~Pontecorvo 3, I-56127 Pisa, Italy.}
\affiliation{INFN, Sezione di Pisa, Largo B.~Pontecorvo 3, I-56127 Pisa, Italy.}
\author{Francesco Negro}\email{francesco.negro@davigonicoloso.edu.it}
\affiliation{INFN,
  Sezione di Pisa, Largo B.~Pontecorvo 3, I-56127 Pisa, Italy.}
\affiliation{
Istituto per l'Istruzione Superiore "G. Da Vigo - N. Da Recco" Via Don 
Giovanni Minzoni 1, I-16035 Rapallo, Italy}
\author{Andrea Rucci}\email{andrea.rucci@pi.infn.it}
\affiliation{Dipartimento di Fisica dell'Universit\`a di Pisa, Largo B.~Pontecorvo 3, I-56127 Pisa, Italy.}
\affiliation{INFN, Sezione di Pisa, Largo B.~Pontecorvo 3, I-56127 Pisa, Italy.}
\author{Francesco Sanfilippo}\email{francesco.sanfilippo@roma3.infn.it}
\affiliation{INFN, Sezione di Roma Tre, Via della Vasca Navale 84, I-00146 Rome, Italy.\\}
\date{\today}

\begin{abstract}
  We study the dependence of the static quark free energy on the
  baryon chemical potential for $N_f = 2+1$ QCD with physical quark
  masses, in a range of temperature spanning from 120~MeV up to 1~GeV
  and adopting a stout staggered discretization with two different
  values of the Euclidean temporal extension, $N_t = 6$ and $N_t = 8$.
  In order to deal with the sign problem, we exploit both Taylor
  expansion and analytic continuation, obtaining consistent results.
  We show that the dependence of the free energy on $\mu_B$ is
  sensitive to the location of the chiral crossover, in
  particular the $\mu_B$-susceptibility, i.e. the linear term in
  $\mu_B^2$ in the Taylor expansion of the free energy, has a peak
  around 150 MeV. We also discuss the behavior expected in the high 
  temperature regime based on perturbation theory, and obtain
  a good quantitative agreement with numerical results.
\end{abstract}

% \pacs{
%   12.38.Gc, %Lattice QCD calculations
%   12.38.Mh, %Quark-gluon plasma
%   12.38.Aw %General properties of QCD (dynamics, confinement, etc.)
% }

\maketitle

\section{Introduction}
\label{intro}

Heavy quark free energies have been used as a probe of the confining
properties of strong interactions since the early days of lattice QCD
simulations. They can be extracted, after proper 
renormalization~\cite{Kaczmarek:2002mc, Petreczky:2004pz, Kaczmarek:2007pb, Kaczmarek:2005gi, Borsanyi:2015yka},
from the expectation value of the Polyakov loop and of its
correlators. The Polyakov loop is defined in the continuum as
\begin{equation}
  \label{eq:polyloop}
  L(\mathbf{r}) =
  \frac{1}{N_c}\mathcal{P}\exp
  \left(ig\int_{0}^{1/T}d\tau A_{0}(\mathbf{r},\tau)\right) ~,
\end{equation}
where $T$ is the temperature, $\mathcal{P}$ is the path-ordering
operator, and $N_c$ is the number of colors. On the lattice, this
object is constructed by taking the product of gauge links winding
along the compactified Euclidean temporal direction. The square module
of its trace is the asymptotic value of the unsubtracted correlator
between Polyakov loops: it is related to the static quark free energy
$F_Q$ by the formula
\begin{equation}
2 F_Q = -{T} \log 
|\left\langle \mathrm{Tr}L \right\rangle|^2 \, .~
\end{equation}

In the pure gauge theory the Polyakov loop is an exact order parameter
for color confinement/deconfinement, which becomes non-zero only in
the deconfined phase and signals the spontaneous breaking of center
symmetry. This is usually associated with the possibility of
separating two static color charges at arbitrarily large distances
without paying an infinite amount of free energy.

In full QCD the situation is different: the creation of dynamical
quark-antiquark pairs makes the free energy of static quark pairs
finite at any distance even in the confined phase. In fact, dynamical
quarks break center symmetry explicitly, so that the Polyakov loop is
not an exact order parameter any more and its expectation value is
different from zero even in the confined phase.

In the presence of physical quark masses chiral symmetry is surely a
relevant symmetry, even if not exact, and the chiral condensate and
its susceptibility are usually adopted as probes to locate the
pseudo-critical temperature of QCD, which is found to be around 
155~MeV~\cite{aefks,afks,betal,tchot,tchot2,tchot3}.  
Still, the Polyakov loop shows a rapid rise at a similar
temperature scale, signalling the passage to a deconfined regime with
screened color interactions.

Whether deconfinement and chiral symmetry restoration take place at
exactly the same temperature is yet not clear and maybe not even a
well founded question. The Polyakov loop susceptibility shows a peak
around 200 MeV~\cite{Bazavov:2016uvm}, while other related observables
show a signal closer to the chiral transition temperature: this is the
case for the Polyakov loop entropy $S_Q = -\partial F_Q / \partial
T$~\cite{Bazavov:2016uvm} or the so-called transverse susceptibility
related to fluctuations in the imaginary part of the Polyakov
loop~\cite{Lo:2013etb,Lo:2013hla}. Since the QCD transition is
actually a crossover, it is quite natural to expect that different
observables yield different locations of the pseudo-critical
temperature. Yet, the information coming from different probes can be
useful to better understand the connection between different phenomena
taking place around the crossover region.

The purpose of the present study is to give a closer look at static
quark free energies, in particular by exploring their dependence on
the baryon chemical potential $\mu_B$.  The modification of the heavy
quark free energy due to $\mu_B$,
$\Delta F_Q(T,\mu_B)\equiv F_Q(T,\mu_B)-F_Q(T,0)$, is given by the
following expression
\begin{equation}
  \label{def:deltafq}
  \frac{\Delta F_Q(T,\mu_B)}{T} = -\log \left( 
    \frac{|\langle \mathrm{Tr}L\rangle (T,\mu_B)|} 
    {|\langle\mathrm{Tr}L\rangle(T,~0~)|} 
  \right) \, ,
\end{equation}
which does not need renormalization if the two Polyakov loops in the
ratio are computed at the same ultraviolet (UV) scale. This quantity
has been studied in Ref.~\cite{Doring:2005ih} and more recently in
Ref.~\cite{Andreoli:2017zie} for QCD with physical quark masses.

One expects the dependence of $\Delta F_Q(T,\mu_B)$ on $\mu_B$ to be
sensitive to the location of the transition. Indeed, if the Polyakov
loop were an exact order parameter then its dependence on $\mu_B$
should become singular at $T_c$, because $\mu_B$ is a relevant
parameter which modifies the location of $T_c$. A remnant of this
behavior must be present even when the Polyakov loop is not an exact
order parameter and, since the free energy is an even function of
$\mu_B$, the first non-trivial derivative to investigate the associate
pseudocritical behavior is the mixed susceptibility
\begin{equation}
  \chi_{Q,\mu_B^2} \equiv
  - \frac{\partial^2 (F_Q/T)}{\partial
    (\mu_B/T)^2}\bigg|_{\mu_B=0}\, .
  \label{def:mixsusc}
\end{equation}
Early simulations of $N_f = 2$ QCD have shown that this quantity has a broad 
peak in a region close to $T_c$~\cite{Doring:2005ih}. More recent simulations,
performed for $N_f = 2+1$ QCD discretized via stout-staggered fermions
with physical quark masses~\cite{Andreoli:2017zie}, were limited
to a temperature range $T~\gtrsim~180$~MeV, showing nevertheless a
peculiar behavior pointing to a
seeming divergence for $T \sim 150$ MeV.

The purpose of the present study is to extend the investigation 
for $N_f = 2+1$ QCD with physical quark masses 
to a wider temperature range, going from 120~MeV up to
1~GeV. We consider the same stout-staggered discretization
adopted in Ref.~\cite{Andreoli:2017zie} and
two different sets of
lattice spacings, corresponding to 
Euclidean temporal extensions $N_t = 6$ and $N_t = 8$,
in order to estimate the impact of systematic errors
related to the UV cutoff.  
The extended range of temperatures will permit us both
to investigate the pseudocritical behavior of 
  $\chi_{Q,\mu_B^2}$ around $T_c$, and to 
compare results obtained at high $T$
with perturbative predictions.
Since lattice simulations at non-zero
$\mu_B$ are not feasible, because of the sign problem, we employ both
Taylor expansion and analytic continuation from simulations at
imaginary $\mu_B$ in order to properly cover the whole temperature
range: for temperatures where both methods are used we obtain
consistent results.

The paper is organized as follows: in Section~\ref{secii} we review
our numerical methods and the observables explored in this study;
results are presented in Section~\ref{res} and, finally, in
Section~\ref{concl}, we draw our conclusions.

\section{Numerical setup and observables}
\label{secii}

We have considered the finite temperature partition function for
${N_f=2+1}$ QCD with chemical potentials $\mu_f$ ($f = u,d,s$) coupled
to quark number operators, ${\mathcal Z}(T,\mu_u,\mu_d,\mu_s)$, in a
setup for which $\mu_u = \mu_d = \mu_s = \mu_B/3$, corresponding to a
purely baryonic chemical potential.  The path integral formulation of
${\mathcal Z} (T,\mu_B)$, discretized via improved rooted staggered
fermions and adopting the standard exponentiated implementation of the
chemical potentials~\cite{Hasenfratz:1983ba,Gavai:1985ie}, reads
\begin{equation}
  \mathcal{Z} = \int \mathcal{D}U e^{- \mathcal{S}_{\text{YM}}}
  \prod_{f=u,d,s}\det\left[ M_{\text{st}}^{f}(U,\mu_{f}) \right]^{1/4} ~, 
\label{partfunc}
\end{equation}
where
\begin{equation}
  \mathcal{S}_{\text{YM}} = -\frac{\beta}{3} \sum_{i,\mu\neq\nu}
  \left( \frac{5}{6}W_{i;\mu\nu}^{1\times1}
    - \frac{1}{12}W_{i;\mu\nu}^{1\times2} \right)
\end{equation}
is the tree-level Symanzik improved action~\cite{Weisz:1982zw,
  Curci:1983an} ($W_{i;\mu\nu}^{n\times m}$ stands for the trace of
the $n\times m$ rectangular parallel transport in the $\mu$-$\nu$
plane and starting from site $i$), and the staggered fermion matrix is
defined as
\begin{eqnarray}
  M_{\text{st}}^{f}(U,\mu_{f})
  & = &
        am_f\delta_{i,j} +
        \sum_{\nu=1}^{4} \frac{\eta_{i;\nu}}{2}\big[
        e^{ a\mu_{f}\delta_{\nu,4}}U_{i;\nu}^{(2)}
        \delta_{i,j-\hat{\nu}}
        \nonumber\label{fermatrix}\\ 
  & - & e^{-a\mu_{f}\delta_{\nu,4}}U_{i-\hat{\nu};\nu}^{(2)\dagger}
        \delta_{i,j+\hat{\nu}}
        \big]~,
\end{eqnarray}
where $U_{i;\nu}^{(2)}$ are two-times stout-smeared links, with
isotropic smearing parameter $\rho=0.15$~\cite{Morningstar:2003gk}.
Bare parameters have been set so as to stay on a line of constant
physics~\cite{Aoki:2009sc, Borsanyi:2010cj, Borsanyi:2013bia}, with
equal light quark masses, $m_u=m_d=m_l$, {a physical 
strange-to-light mass
ratio, $m_s/m_l=28.15$, and a physical pseudo-Goldstone pion mass, 
$m_{\pi}\simeq135~\text{MeV}$.}

The main observable we are interested in is the Polyakov loop and its
dependence on $\mu_B$.  In particular, as already described above, the
ratio of Polyakov loops at different baryon chemical potentials gives
access to the $\mu_B$-dependent part of the free energy density,
$\Delta F_Q(T,\mu_B) \equiv F_Q(T,\mu_B) - F_Q(T,0)$,
\begin{equation}
  \label{def:deltafq}
  \frac{\Delta F_Q(T,\mu_B,\beta)}{T} = -\log \left( 
    \frac{|\langle \mathrm{Tr}L\rangle (T,\mu_B,\beta)|} 
    {|\langle\mathrm{Tr}L\rangle(T,~0~,\beta)|} 
  \right)~,
\end{equation}
and if the ratio is taken for Polyakov loops measured at the same
value of the inverse bare coupling $\beta$ and of the bare quark masses, 
then no further
renormalization is expected, at least when the chemical potential is
inserted on the lattice with the prescription introduced in
Ref.~\cite{Hasenfratz:1983ba} and adopted in the present
investigation.  That means that the dependence of
$\Delta F_Q(T,\mu_B,\beta)$ on $\beta$ is expected to be limited to
finite UV corrections to continuum scaling.

It would be interesting to study the dependence of $F_Q$ on $\mu_B$ in
the whole range of physically relevant values of $\mu_B$, however our
investigation will be limited to the region of small $\mu_B/T$ and, in
particular, to the susceptibility $\chi_{Q,\mu_B^2}$ defined in
Eq.~(\ref{def:mixsusc}), which can be directly related to the Polyakov
loop ratio of Eq.~(\ref{def:deltafq}) by the formula
\begin{equation}
  \frac{|\langle \mathrm{Tr}L\rangle(T,\mu_B)|} 
  {|\langle \mathrm{Tr}L\rangle (T,~0~)|} =   
  1 + \frac{1}{2} \chi_{Q,\mu_B^2} \left(\frac{\mu_B}{T}\right)^2 
  + \mathcal{O}\left(\left(\frac{\mu_B}{T}\right)^4\right)
  \label{eq:ratioloopfit}
\end{equation}
since, from Eq.~(\ref{def:deltafq}), one has
\begin{equation}
  \frac{\partial^2}{\partial(\mu_B/T)^2}
    \frac{|\langle\textnormal{Tr}L\rangle(T,\mu_B)|}
    {|\langle\textnormal{Tr}L\rangle(T,~0~)|}\bigg|_{\mu_B=0} \hspace{-8pt}
= - \frac{\partial^2 (F_Q/T)}{\partial
      (\mu_B/T)^2}\bigg|_{\mu_B=0} \hspace{-8pt} 
    \label{def:curvature}
\end{equation}
The reason of the limitation to small chemical potentials is the well
known sign problem of QCD at finite density, which makes standard
Monte-Carlo simulations unfeasible when $\mu_B \neq 0$. Present
strategies to partially circumvent the sign problem are reliable only
in a limited range of small $\mu_B/T$, where they lead to controllable
systematic errors; Taylor expansion~\cite{tay1,tay2,tay3,tay4} and
analytic continuation from simulations at imaginary chemical
potential~\cite{alford,lomb99,fp1,dl1,azcoiti,chen,Wu:2006su,NN2011,
  giudice,ddl07,cea2009,alexandru,cea2012,Karbstein:2006er,cea_other,
  sanfo1, Takaishi:2010kc,cea_hisq1,corvo,nf2BFEPS,bellwied,gunther,
  gagliardi,Bornyakov:2017upg} are the most widely used techniques. In
this investigation we employ both of them, since in part of our wide
temperature range the statistical or systematic errors of one
technique are less under control, so that a direct comparison with the
other technique improves the overall reliability of the results; this
combined strategy has revealed successful in other cases, like for the
determinations of the curvature of the pseudo-critical
line~\cite{Bonati:2018nut}.

In the analytic continuation approach, the baryon chemical potential
is taken to be purely imaginary, $\mu_B=i\mu_{B,I}$, the path-integral
measure staying real and positive for $\mu_{B,I} \neq 0$.  Within
our numerical setup, adding a non-zero $\mu_{B,I}$ can be rephrased in
terms of a rotation of temporal boundary conditions of the quark
fields by a factor $\exp(i \mu_I /T)$, where $\mu_I=\mu_{B,I}/3$ is
the imaginary part of the quark chemical potential.  
The value of the Polyakov loop is measured for
several values of $\mu_I$ at fixed temperature, then numerical data
are fitted to the analytic continuation of some suitable ansatz for
the dependence on $\mu_B$, thus fixing the corresponding parameters.
Despite its simplicity, this method has some limitations and
drawbacks, its systematic errors being related essentially to the
arbitrary ansatz for the fitting function.

\begin{figure}
  \includegraphics*[width=\columnwidth]{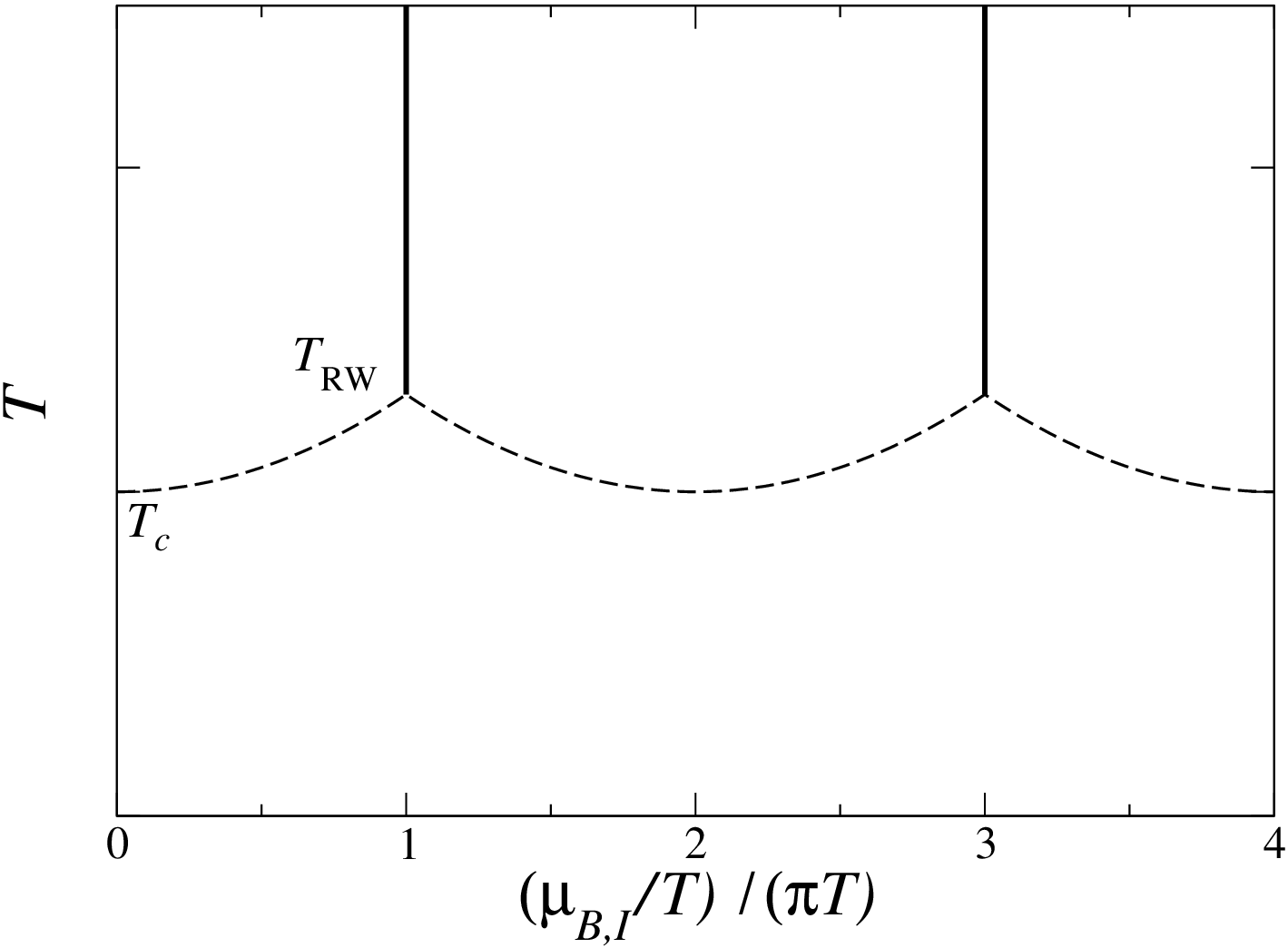}
  \caption{Qualitative structure 
of the QCD phase diagram of QCD in the $T - \mu_{B,I}$
plane. The vertical lines are the RW transitions, while the 
dashed line is the analytic 
continuation of the pseudo-critical line.}
  \label{fig:immu_diag}
\end{figure}

The choice of the fitting function and the related systematics
can be different depending on the value of the temperature, 
as dictated by the non-trivial symmetries and phase structure
of the $T - \mu_{B,I}$ phase diagram, which
is sketched in Fig.~\ref{fig:immu_diag}. In general one can prove,
combining $\mu_{B,I}$ translations with gauge field
center transformations, that the theory is $2 \pi$-periodic
in $\mu_{B,I}/T$~\cite{rwpaper}. This periodicity is smoothly realized
for $T < T_c$: there a Fourier expansion is the most natural choice~\cite{dl1}
and, moreover, a picture based on the Hadron Resonance Gas (HRG) model 
suggests an ansatz where the first few terms of the expansion
are dominant, unless one is close enough to $T_c$.

On the contrary, at high $T$, in particular 
for $T > T_{RW}$ 
(where $T_{RW} \simeq 210$~MeV in the continuum limit
for $N_f = 2+1$ QCD with physical quark masses~\cite{Bonati:2016pwz}),
the periodicity is realized
in a non-analytic way, with first order 
phase transition lines (RW-lines) crossed 
for $\mu_{B,I}/T = (2 k + 1) \pi$ and $k$ integer: the 
phase of the Polyakov loop is an order parameter for such 
transitions, at which the systems switches from 
one center sector to the other.
%; its modulus is however PERCHE HOWEVER? PERCHE QUESTA FRASE, PIU CHE ALTRO?
%continuous there, with a cusp in its first derivative.
% [REFS].
That limits the range of chemical potentials 
available for analytic continuation
to $\mu_{B,I}/T < \pi$, however the dependence of the 
Polyakov loop modulus is 
well approximated by an even power law expansion in 
$\mu_{B,I}$, with the lowest order terms becoming 
more and more dominant as the temperature is increased.

The intermediate region, $T_c < T < T_{RW}$, is the one where
systematic errors can be more severe. In this region, moving in
$\mu_{B,I}/T$ from 0 to $\pi$ one crosses the analytic continuation of
the pseudocritical line: even if this is not a true transition but
just a crossover, it can make the dependence on $\mu_{B,I}$
non-trivial, thus in fact restricting the region of $\mu_{B,I}$ where
different ansatzs give consistent results; moreover, such a region is
smaller and smaller as $T_c$ is approached from above.

A second possibility, which can be put in the 
general framework of the Taylor expansion approach, 
is to measure 
$\chi_{Q,\mu_B^2}$ directly 
 at $\mu_B = 0$, following its definition 
in Eq.~(\ref{def:mixsusc}). In particular, after some computations 
(which are reported in the appendix), one writes
$\chi_{Q,\mu_B^2}$ as a combination of correlators 
involving the Polyakov loop and fermionic terms. The
expression is
\begin{eqnarray}
  \chi_{Q,\mu_B^2}
  &=& 
      \frac{\langle\textnormal{ReTr}L(n^2+n')\rangle}
      {\langle\textnormal{ReTr}L\rangle} -
      \langle n^2+n'\rangle \nonumber\\
  &+& \frac{\langle(\textnormal{ReTr}L+\textnormal{ImTr}L)n\rangle^2}
      {\langle\textnormal{ReTrL}\rangle^2}~,
  \label{def:curvature2}
\end{eqnarray}
where $n=n_u+n_d+n_s$ is the total quark number and $n'$ is its derivative
with respect to $\mu_B$. Even though the measure of this
quantity is well defined and seemingly straightforward for all temperatures,
in practice its computation involves many noisy estimators 
and therefore turns out to be numerically expensive, 
especially in the region around and below $T_c$.

In view of the above considerations, 
the strategy chosen in this work has been 
to adopt analytic continuation for all
temperatures below $T_c$ and for most temperatures above $T_{RW}$,
while in the region $T_c < T < T_{RW}$ we 
have adopted both Taylor expansion and analytic continuation,
in order to have better control over systematics.
\\

Monte-Carlo simulations have been performed for two 
different values of $N_t$ in order to estimate 
the impact of UV corrections, in particular on a ${24^3\times 6}$ and
on a ${32^3\times 8}$ lattices using a Rational Hybrid Monte-Carlo
algorithm~\cite{rhmc1, rhmc2, rhmc3}.
A summary of the parameters adopted in our simulations, together
with details on the strategy chosen in each case, is
reported in Tab.~\ref{tab:betas}.
\begin{table}[t!]
  \centering
  \label{tab:betas}
  \begin{tabular}{ c c c c c }
    \hline\hline
    $N^3\times N_t$ & $\beta$ & $a~[\mathrm{fm}]$
    & $T~[\mathrm{MeV}]$ & $\mu_I/(\pi T)$\\
    \hline\\[-1.0em]
    $24^3\times6$ & 3.4500 & 0.2835 & 116  & $0,0.04,\dots,0.32$ \\
    ''            & 3.4789 & 0.2631 & 125  & $0,0.04,\dots,0.32$ \\
    ''            & 3.5085 & 0.2436 & 135  & $0,0.04,\dots,0.32$ \\
    ''            & 3.5246 & 0.2332 & 141  & $0,0.04,\dots,0.64$ \\
    ''            & 3.5421 & 0.2222 & 148  & $0$ \\
    ''            & 3.5585 & 0.2121 & 155  & $0$ \\
    ''            & 3.5695 & 0.2055 & 160  & $0$ \\
    ''            & 3.5800 & 0.1993 & 165  & $0,0.04,\dots,0.32$ \\
    ''            & 3.5923 & 0.1923 & 171  & $0$ \\
    ''            & 3.6172 & 0.1787 & 184  & $0,0.04,\dots,0.32$ \\
    ''            & 3.6746 & 0.1515 & 217  & $0,0.04,\dots,0.32$ \\
    ''            & 3.7305 & 0.1310 & 251  & $0,0.04,\dots,0.32$ \\
    ''            & 3.7829 & 0.1153 & 285  & $0,0.04,\dots,0.32$ \\
    ''            & 3.8300 & 0.1034 & 318  & $0,0.04,\dots,0.32$ \\
    ''            & 3.8749 & 0.0936 & 351  & $0,0.04,\dots,0.32$ \\
    ''            & 3.9184 & 0.0856 & 384  & $0,0.04,\dots,0.32$ \\
    ''            & 3.9608 & 0.0788 & 417  & $0,0.04,\dots,0.32$ \\
    ''            & 4.0019 & 0.0729 & 451  & $0,0.04,\dots,0.32$ \\
    ''            & 4.0798 & 0.0635 & 518  & $0$ \\
    ''            & 4.1506 & 0.5622 & 585  & $0$ \\
    ''            & 4.2200 & 0.0504 & 652  & $0$ \\
    ''            & 4.2797 & 0.4574 & 719  & $0$ \\
    ''            & 4.3297 & 0.0418 & 786  & $0$ \\
    ''            & 4.3778 & 0.0386 & 853  & $0$ \\
    ''            & 4.4284 & 0.0357 & 920  & $0$ \\
    ''            & 4.4808 & 0.0333 & 987  & $0$ \\
    ''            & 4.5317 & 0.0312 & 1054 & $0$ \\
    ''            & 4.5764 & 0.0293 & 1121 & $0$ \\
    \hline\\[-1.0em]
    $32^3\times8$ & 3.5835 & 0.1973 & 125  & $0,0.04,\dots,0.32$ \\
    ''            & 3.6100 & 0.1827 & 135  & $0,0.04,\dots,0.64$ \\
    ''            & 3.6245 & 0.1749 & 141  & $0,0.04,\dots,0.64$ \\
    ''            & 3.6417 & 0.1666 & 148  & $0,0.04,\dots,0.64$ \\
    ''            & 3.6570 & 0.1591 & 155  & $0$ \\
    ''            & 3.6700 & 0.1541 & 160  & $0,0.02,\dots,0.16$ \\
    ''            & 3.6800 & 0.1494 & 165  & $0,0.02,\dots,0.24$ \\
    ''            & 3.6925 & 0.1442 & 171  & $0,0.04,\dots,0.32$ \\
    ''            & 3.7250 & 0.1333 & 185  & $0$ \\
    ''            & 3.8525 & 0.0982 & 251  & $0$ \\
    ''            & 4.1678 & 0.0546 & 451  & $0,0.04,\dots,0.32$ \\
    ''            & 4.2560 & 0.0476 & 518  & $0$ \\
    ''            & 4.3255 & 0.0422 & 585  & $0$ \\
    ''            & 4.3899 & 0.0378 & 652  & $0$ \\
    ''            & 4.4586 & 0.0343 & 719  & $0$ \\
    ''            & 4.5273 & 0.0314 & 786  & $0$ \\
    ''            & 4.5861 & 0.0289 & 853  & $0$ \\
    \hline\hline
  \end{tabular}
  \caption{List of parameters used in the Monte-Carlo simulations for
    the study of the susceptibility $\chi_{Q,\mu_B^2}$, chosen so as
    to stay on a line of constant physics at the physical point, using
    a spline interpolation of the data in Refs.~\cite{Borsanyi:2010cj,
      Borsanyi:2013bia}.}
\end{table}
In the cases in which the susceptibility $\chi_{Q,\mu_B^2}$ has been
measured through Taylor expansion, sets of about $10^4$ configurations
separated by 10 molecular dynamics trajectories have been analyzed for
each run, and fermionic observables such as the quark number $n$ and
its derivative $n'$ have been computed through {stochastic noisy 
estimators~\cite{Dong:1993pk}, in particular using up to 256 $Z_2$
random noise vectors per measurement.}
In the cases
in which analytic continuation has been adopted, we have  
performed around $5\times10^3$ molecular dynamics 
trajectories for each value of the imaginary
chemical potentials. The data analysis has been performed 
by means of a blocked jackknife resampling in all cases.

\section{Results}\label{res}

\begin{figure}
  \includegraphics*[width=\columnwidth]{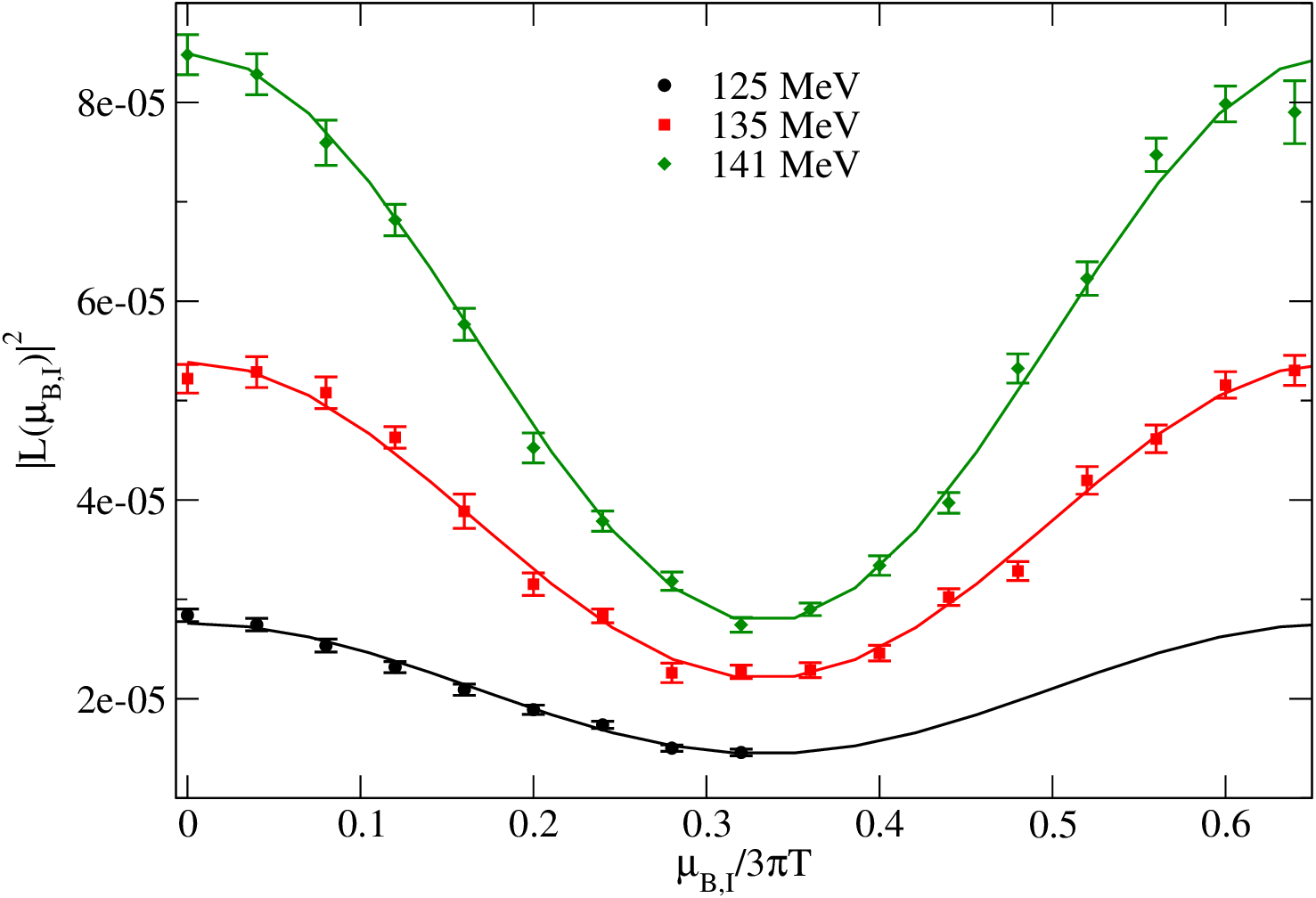}\\~\\
  \includegraphics*[width=\columnwidth]{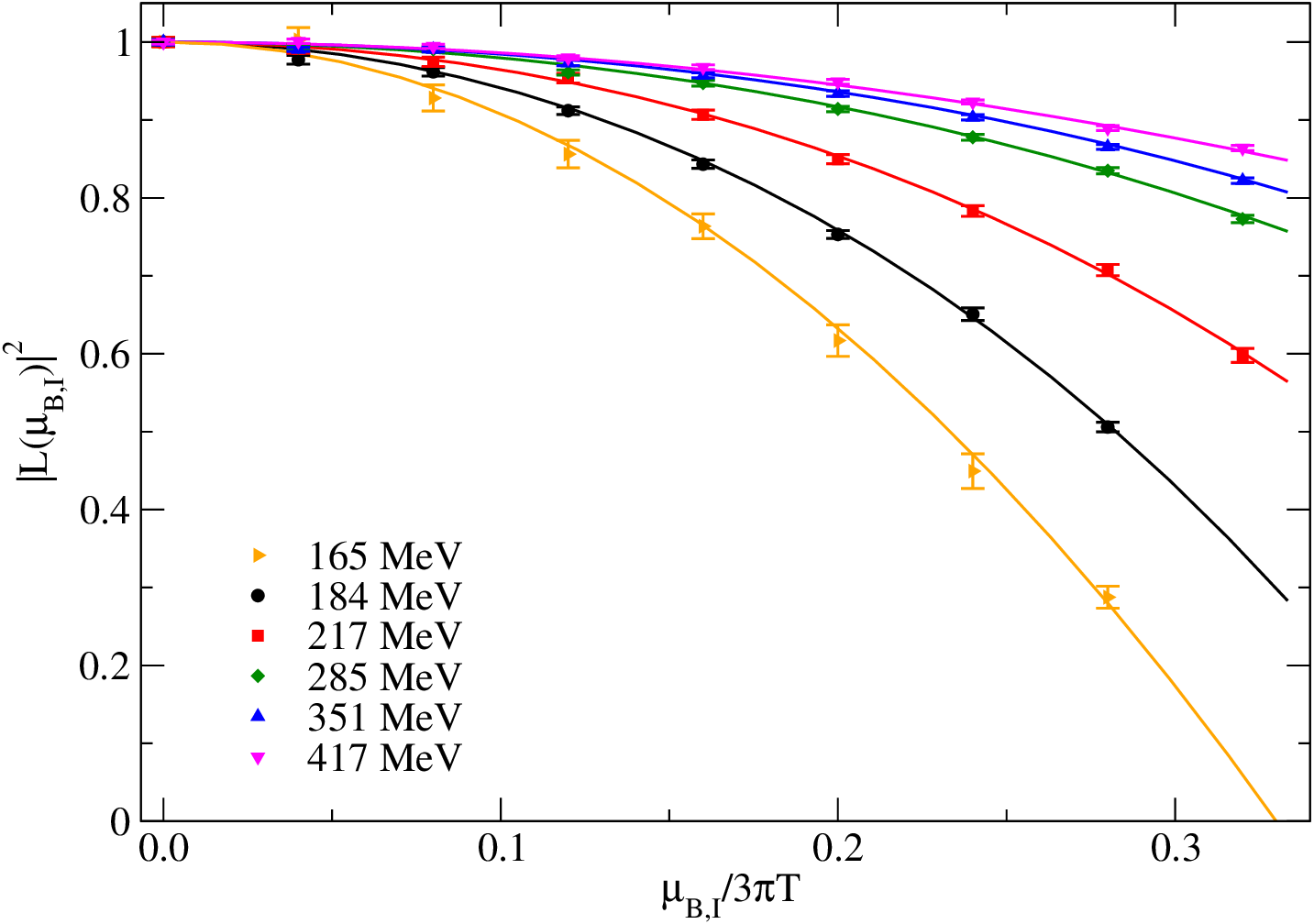}
  \caption{Square module of the Polyakov loop as a function of the
    imaginary chemical for several temperatures below (top) and above
    (bottom, normalized to the value at $\mu=0$) the pseudo-critical
    temperature $T_c\simeq155$~MeV \cite{corvo}, measured on the
    $24^3\times6$ lattice. Curves are the results of the fit using,
    respectively, the cosine expansion in Eq.~\eqref{eq:chifit_cosine}
    and the polynomial ansatz in Eq.~\eqref{eq:highT_exp}.}
  \label{fig:polyloop_res}
\end{figure}

Let us start by discussing the 
determination of $\chi_{Q,\mu_B^2}$ by analytic continuation.
As an illustrative example, 
in Fig.~\ref{fig:polyloop_res} we report the average values 
of the squared modulus of the Polyakov loop
on the $24^3 \times 6$ lattice as a function of $\mu_{B,I}$ and 
for some of the explored temperatures. For the sake of readability,
we have reported separately determinations at high and low $T$, normalizing
data by the value at $\mu_{B,I} = 0$ only in the latter case.

At low temperatures, as a matter of fact, we have found
that a single cosine
term is sufficient to correctly describe our data
for all explored temperatures, i.e.~with values 
of the $\chi^2/{\rm d.o.f.}$ regression parameter close to one:
\begin{equation}
  \frac{|\langle L\rangle(\mu_{B,I})|^2}{|\langle L\rangle(0)|^2}
  = 1 - 2\, \chi_{Q,\mu_B^2}\left[1-\cos\left(\frac{\mu_{B,I}}{T}\right)\right] \, .
  \label{eq:chifit_cosine}
\end{equation}
This allows to determine $\chi_{Q,\mu_B^2}$. We have considered in the
final error also the variability which is obtained by adding a further
term in the Fourier expansion\footnote{ Notice that the
  parametrization in Eq.~(\ref{eq:chifit_cosine}) changes if other
  Fourier terms are added, since in this case $\chi_{Q,\mu_B^2}$ takes
  contributions from all Fourier coefficients.},~i.e. a term
proportional to $\cos({ 2\, \mu_{B,I}}/{T})$.

In the high-temperature regime, instead, we have adopted
a polynomial expansion truncated to the quartic term in 
$\mu_{B,I}$,~i.e.
\begin{equation}
  \frac{|\langle L\rangle(\mu_{B,I})|^2}{|\langle L\rangle(0)|^2}
  = 1 - \chi_{Q,\mu_B^2} \left(\frac{\mu_{B,I}}{T}\right)^2 + l_4\, 
\left(\frac{\mu_{B,I}}{T}\right)^4 \, .
  \label{eq:highT_exp}
\end{equation}
In all cases the fit range has been limited by the location 
of the pseudocritical value of $\mu_{B,I}$ for the given
temperature, as extracted from data reported in 
Refs.~\cite{corvo}, and appropriate systematic uncertainties
have been added to the fit parameters, which take into
account the variability under changes of the fitted range.
We have found that the quartic coefficient $l_4$ is not needed
to obtain reasonable fits (and 
turns out to be compatible with zero when included)
for temperatures $T > T_{RW}$, while for lower temperatures
it is definitely needed in order to get
$\chi^2/{\rm d.o.f.} \sim 1$.

In the region above $T_c$, where the pseudo-critical behavior is more
pronounced, and in some cases also for the same temperatures at which
analytic continuation has been used, we adopted the Taylor expansion
method, measuring directly the value $\chi_{Q,\mu_B}$ through the
formula in Eq.~\eqref{def:curvature2}. The computation, especially
close to $T_c$, turned out to be numerically expensive and, in
general, the uncertainties associated to the measures obtained by this
method are larger than those extracted by analytic
continuation. Nevertheless, in this way no source of systematics is
present and, at least at our level of precision, the estimations make
the picture clear enough. Moreover, for the temperatures where both
methods are available, a reasonable agreement is observed.

\begin{figure}
  \includegraphics*[width=\columnwidth]{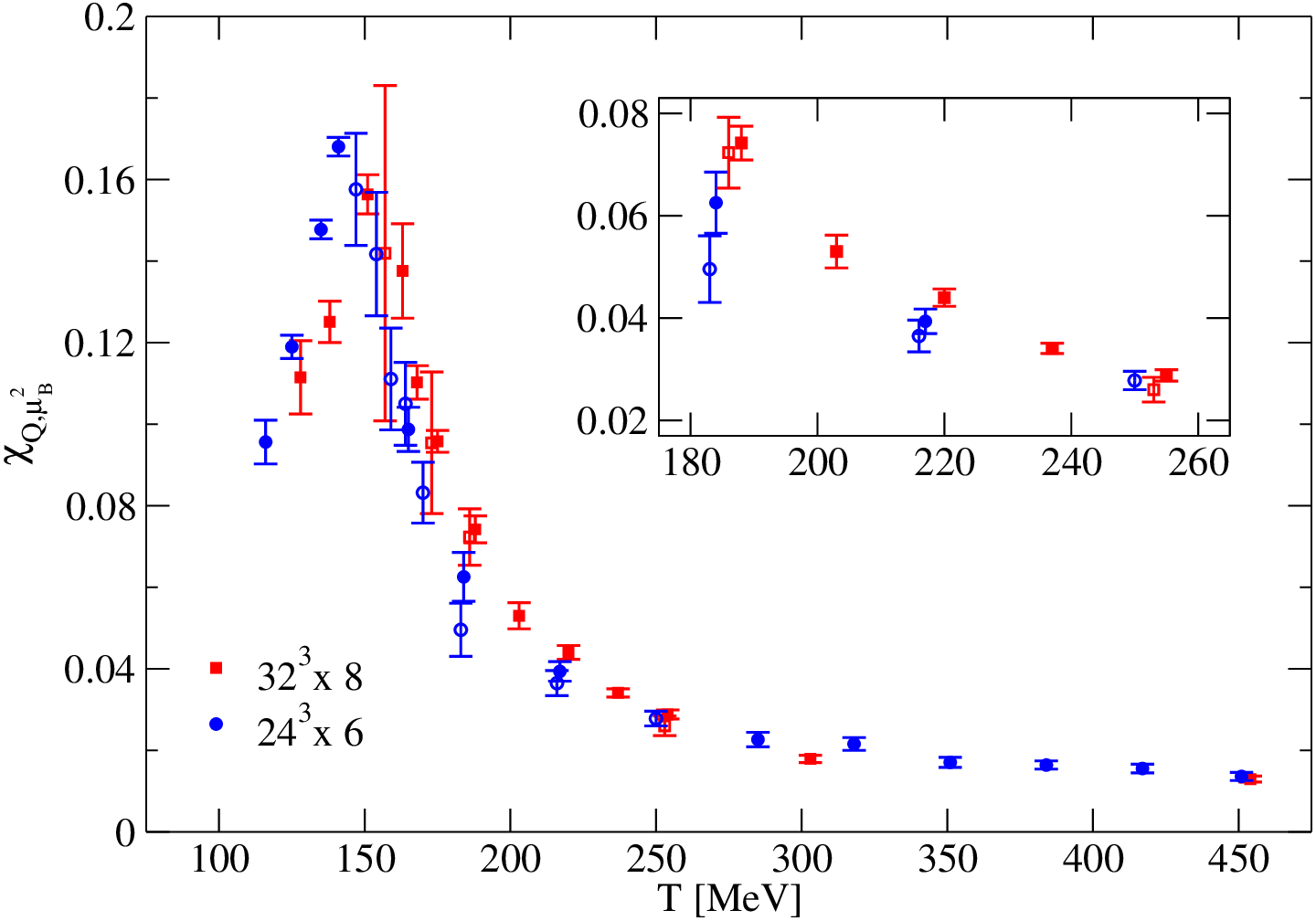}
  \caption{Susceptibility $\chi_{Q,\mu_B^2}$ as function of the
    temperature $T$ extracted from two different lattices
    $24^3\times6$ and $32^3\times8$. The pattern of the dots indicates
    the method used for the computation, with empty and full
    datapoints corresponding, respectively, to the Taylor expansion
    method and to analytic continuation. For some values of the
    temperature, see e.g. the inset, both procedures have been used,
    so as to check the consistency of the
    results. Datapoints have been slightly shifted for the sake of
    readability.} \label{fig:poly_curvature}
\end{figure}

\begin{figure}
  \includegraphics*[width=\columnwidth]{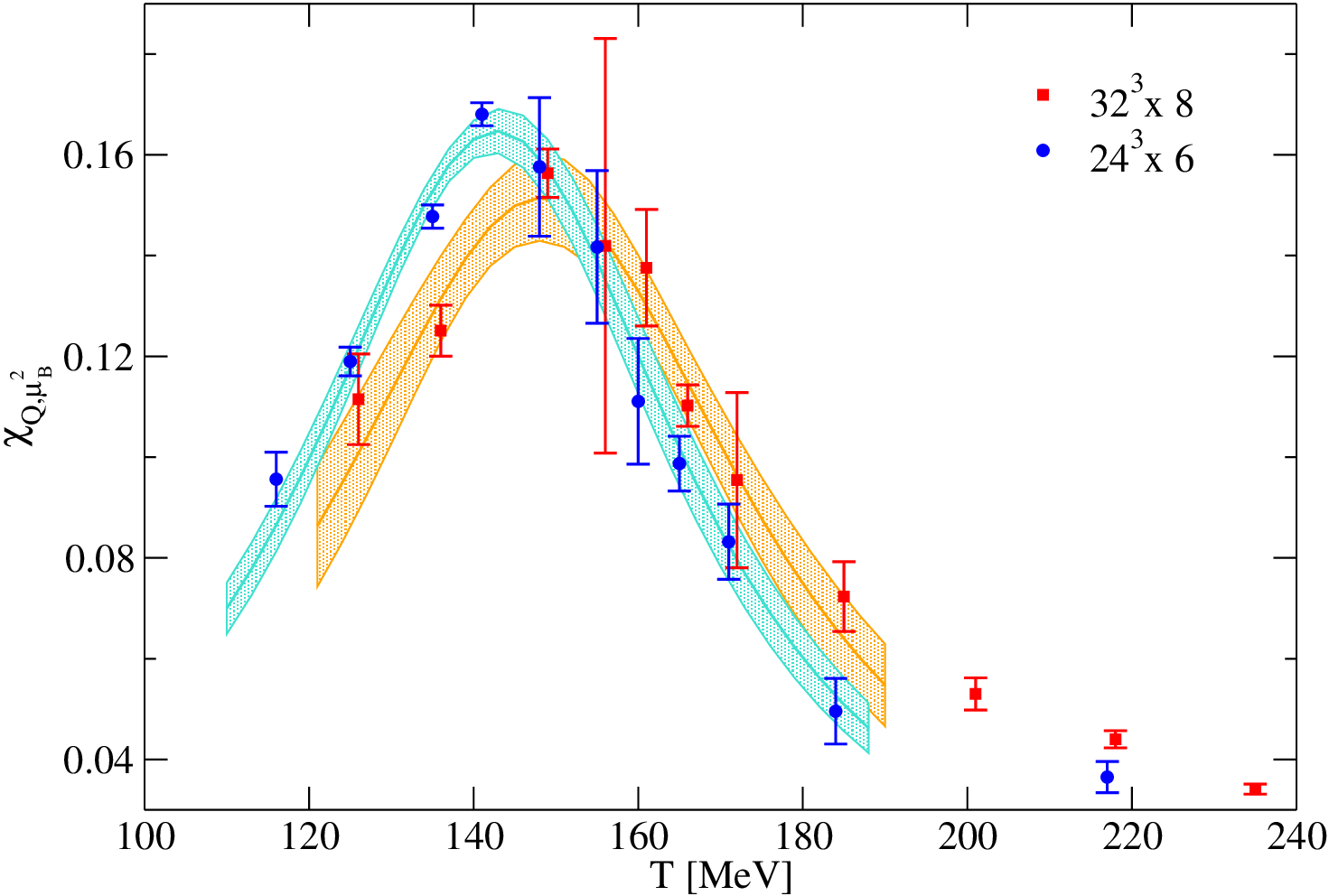}
  \caption{$\chi_{Q,\mu_B^2}$ as a function of $T$ in the region
    near the peak. Curves are the result of best fits to the Lorentzian form
    in Eq.~\eqref{eq:chifit_peak}, where bands are the 68\% CIs
    plotted over the fit range. Reasonable values of 
    $\tilde \chi^2$ have been obtained for both datasets: 
    $\chi^2/{\rm d.o.f.} = 12.4/7$ and $\chi^2/{\rm d.o.f.} = 7.2/5$
    respectively for the 
    the $24^3\times6$ and the
    $32^3\times8$ lattice.}
  \label{fig:poly_curvature_fit}
\end{figure}

The whole collection of results, including 
all temperatures and both sets of lattice spacings,
$N_t = 6$ and $N_t = 8$, is reported 
in Fig.~\ref{fig:poly_curvature}.
The dependence on $N_t$ appears to be small, confirming that, even if
no continuum extrapolation is performed in this study, finite UV
cutoff corrections are not large. The susceptibitliy $\chi_{Q,\mu_B}$
grows rapidly in the crossover region near $T_c$, where it exhibits a
well-defined peak.  The location of the peak can be determined
quantitatively by modelling the observed behavior near the maximum. In
particular, we have adopted a Lorentzian function, defined as
\begin{equation}
  \chi_{Q,\mu_B^2} = \frac{p_0}{1+\left[(T-T_{L})/p_1\right]^2} \, , 
  \label{eq:chifit_peak}  
\end{equation}
where $T_{L}$ indicates the pseudo-critical temperature related to
the observable $\chi_{Q,\mu_B^2}$. This ansatz
well describes the peak structure for both values
of $N_t$: best-fit curves are shown in Fig.~\ref{fig:poly_curvature_fit}
and yield $T_L=143.4\pm1.2~\textnormal{MeV}$ and
$T_L=147.7\pm1.4~\textnormal{MeV}$ respectively for $N_t = 6$ and
$N_t = 8$.  The uncertainties include systematics related to the
choice of the fit range, but not those associated with the
determination of the lattice spacing, which are of the order of
$2-3\%$~\cite{Borsanyi:2010cj, Borsanyi:2013bia}.  Similar results are
obtained using a different fitting ansatz, like a purely quadratic
function of $T$.  The small $N_t$-dependence observed for $T_L$ points
to a continuum limit around 150~MeV, which is very close to
$T_c \simeq 155$~MeV.

\section{Comparison with perturbation theory}

Finally, it is interesting to discuss the fate of 
$\chi_{Q,\mu_B^2}$ 
 in the large $T$
limit. At zero baryon chemical potential, $F_{Q}(T)$ is expected to
decrease unboundedly as $T$ increases, a well-known behavior
predicted by weak-coupling calculations
\cite{Gava:1981qd,Berwein:2015ayt} and observed also on the lattice in
many studies \cite{Doring:2005ih, Bazavov:2013yv, Borsanyi:2015yka,
  Bazavov:2016uvm}. At leading order, its expression in the high
temperature regime is given by
\begin{equation}
  \label{eq:pertfq}
  F_{Q}(T)=-\frac{C_F}{2}\frac{g^2}{4\pi}m_D(T)~,
\end{equation}
where $C_F=(N_c^2-1)/2N_c$ is the Casimir operator in the foundamental
representation and $m_D(T)$ is the Debye screening mass
which, at the leading order is
\begin{equation}
  m_D^2(T) = \frac{1}{3}\left(N_c+\frac{N_f}{2}\right)g^2T^2~.
\end{equation}
In the dense medium, screening effects are amplified and the value of
the single quark free energy grows indefinitely (in module). In the
very large temperature limit, at leading order, the expression of
$F_{Q}(T,\mu_B)$ is obtained performing an expansion of the Debye mass
for small values of the chemical potential
\cite{Doring:2005ih,Kaczmarek:2007pb}. The result is the appearance
of a quadratic dependence on $\mu_B$, 
{$F_{Q}(T,\mu_B) = F_{Q}(T)\, m_D(T,\mu_B) / m_D(T)$ where}
\begin{equation}
  m_D^2(T,\mu_B) = m_D^2(T)
  \left[1+\frac{3N_f}{2N_c+N_f}\left(\frac{\mu_B}{3 \pi T}\right)^2\right]~,
\end{equation}
{Inserting this
expression} 
in Eq.~(\ref{def:mixsusc})
one
finds
\begin{eqnarray}
\label{eq:chi_asymptotic}
  \chi_{Q,\mu_B^2}\big|_{T\to\infty}
  &=& -\frac{F_Q(T)}{T}\frac{\partial^2}{\partial(\mu_B/T)^2}\frac{m_D(T,\mu_B)}{m_D(T)}\bigg|_{\mu_B=0} \nonumber \\
  %&= -\frac{N_f}{3(2N_c+N_f)\pi^2}\frac{F_Q(T)}{T}\\
  &=& \frac{C_Fg^3}{24\pi^3}\frac{N_f}{2N_c+N_f}\sqrt{\frac{N_c}{3}+\frac{N_f}{6}} \, .
\end{eqnarray}
Consequently, since the coupling runs to
zero at large $T$, 
%scale \cite{Kaczmarek:2004gv}, 
the susceptibility
$\chi_{Q,\mu_B^2}$ vanishes asymptotically as $g^3$. This means that, in this
regime, {a finite baryon density does not affect the in-medium 
static quark free energy}, its contribution being overrided by the
thermal fluctuations. Notice that the same proportionality to $g^3$
at high $T$ is shown also by static quark entropy
$S_Q=-\partial F_Q/\partial T$ which, asymptotically, is expected to
behave as $S_Q\sim- F_Q/T$ \cite{Berwein:2015ayt,Bazavov:2016uvm}, in
agreement with our calculation.

\begin{figure}
  \includegraphics*[width=\columnwidth]{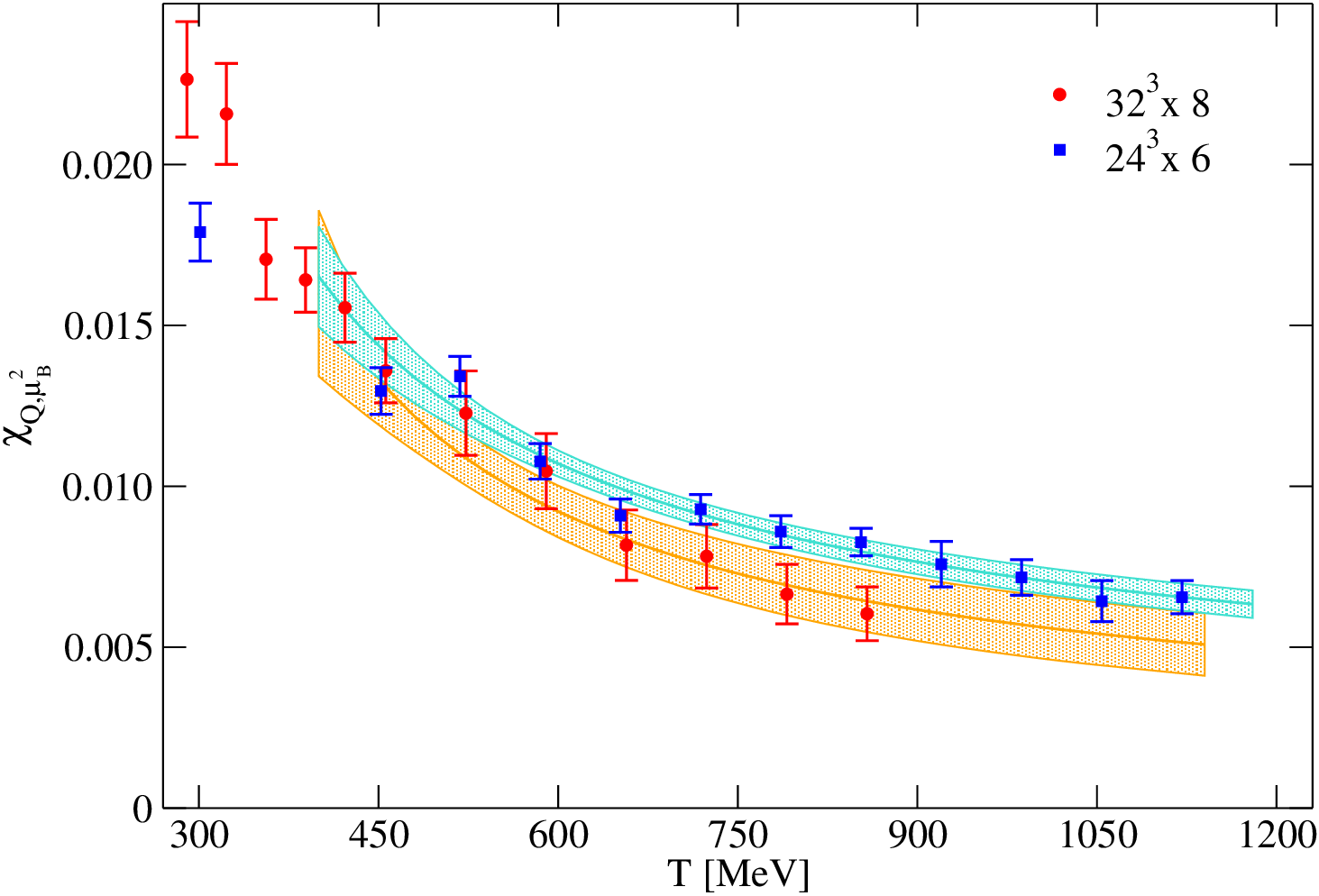}
  \caption{Values of $\chi_{Q,\mu_B^2}$ in the high temperature regime.
    Curves represent best fits to
    Eq.~\eqref{eq:poly_curvature_asymptotic_model}, while bands are
    confidence intervals at 68\% CL. The value of the
    $\chi^2/{\rm d.o.f.}$ test is  $0.5$ and $0.7$
    respectively for the $24^3 \times 6$ and the $32^3 \times 8$ lattice. 
}
  \label{fig:poly_curvature_asymptotic}
\end{figure}

In order to check the consistency of these predictions with lattice
results, we have extended the computation of 
$\chi_{Q,\mu_B^2}$ to higher temperatures, adopting 
the Taylor expansion method which in this regime is not particularly
expensive.
Results are shown in Fig.~\ref{fig:poly_curvature_asymptotic}.
In order to obtain a quantitative prediction from 
Eq.~\eqref{eq:chi_asymptotic}, we need to insert the dependence 
of the coupling constant $g(T)$ on the temperature, which 
at the leading 
order in perturbation theory is given by~\cite{Kaczmarek:2004gv}
\begin{equation}
  g^{-2}(T) = 2\beta_0\log\frac{2\pi T}{\Lambda}
  \quad \beta_0=\frac{11N_f-2N_c}{48\pi^2}~,
\end{equation}
where $\beta_0$ is the first coefficient of the QCD $\beta$-function,
which is independent of the renormalization scheme. Inserting this
expression in Eq.~\eqref{eq:chi_asymptotic} one obtains
\begin{equation}
  \label{eq:poly_curvature_asymptotic_model}
  \chi_{Q,\mu_B^2}\big|_{T\to\infty}
  = p_0\left[\log\frac{2\pi T}{\Lambda}\right]^{-3/2}~,
\end{equation}
where $p_0$ is a pre-factor which is independent of the
renormalization scheme and whose value is $p_0\sim0.019$ 
in our case,
where $N_c=3$ and $N_f=3$ (we assume that
the three quark flavors can be considered
as practically degenerate in this temperature regime).
The slow decrease shown by
the lattice data is well described, both for $N_t = 6$ and 
$N_t = 8$, by Eq.~(\ref{eq:poly_curvature_asymptotic_model}),
the fitted value of $p_0$  being
$0.021(2)$ and $0.014(3)$, respectively, for the $24^3\times6$ 
($\chi^2/{\rm d.o.f.} = 0.73$) and the
$32^3\times8$ ($\chi^2/{\rm d.o.f.} = 0.46$) lattice: we consider such an agreement more than satisfactory,
given that only the leading order has been considered; {it is interesting to 
notice that also the values obtained for the $\Lambda$ parameter 
are reasonable and of the order of 100 MeV.}

\section{Conclusions}
\label{concl}

In this study we have investigated the dependence of the static quark 
free energy on the baryon chemical potential in a wide
temperature range, considering in particular
the leading order dependence, which is quadratic in $\mu_B$
and that we have parameterized in terms of the susceptibility 
$\chi_{Q,\mu_B^2}$. 
The investigation has been carried out by lattice simulations
of $N_f = 2+1$ QCD discretized via stout-staggered fermions
with physical quark masses. Both analytic continuation and Taylor
expansion have been adopted to avoid the sign problem at non-zero
$\mu_B$, obtaining consistent results.

Results for $\chi_{Q,\mu_B^2}$ have been found to be compatible, in
the high temperature regime, with predictions obtained in perturbation
theory. The dependence of the static quark free energy on $\mu_B$
which vanishes as a power law in the gauge coupling $g(T)$, precisely
as $g^3$,~i.e. logarithmically with the temperature $T$. Numerical
results are consistent both with the power law behavior in $g$ and
with the predicted prefactor.

At low temperatures $\chi_{Q,\mu_B^2}$ presents instead a well
defined peak located around 150~MeV, i.e.~roughly compatible
with the crossover temperature $T_c$ corresponding to
the restoration of chiral symmetry. If the Polyakov loop
were an exact order parameter for the deconfinement transition,
one would expect a singular behavior for $\chi_{Q,\mu_B^2}$
at the critical temperature.
Therefore,
the rough coincidence of the two temperatures points 
once again to a strong connection
between chiral symmetry and deconfinement dynamics, even within 
a crossover scenario.

Our results have been obtained for just two sets of lattice spacings,
corresponding to $N_t = 6$ and $N_t = 8$. Future studies should 
extend the investigation to larger values of $N_t$ so as to achieve
a continuum extrapolation for $\chi_{Q,\mu_B^2}$.
However, present results show only 
modest changes as $N_t$ is changed from 6 to 8, so that no 
significant modifications of our conclusions are expected
in the continuum limit.

\acknowledgments
Numerical simulations have been performed on the
MARCONI machine at CINECA, based on the agreement between INFN and
CINECA (under project INF18\_npqcd), at the Scientific Computing
Center at INFN-PISA, {and on the COKA cluster at the University of
  Ferrara and INFN-Ferrara based on the GPU code developed in}
Refs.~\cite{incardona,ferrarapisa,ferrarapisa2}.  FN acknowledges
financial support from the INFN HPC\_HTC project.

\appendix*
\section{Computation of $\chi_{Q,\mu_B^2}$}
\label{appendixchi}
The expression of the curvature $\chi_{Q,\mu_B^2}$ is obtained by
computing the second derivative of the ratio between square modules of
the Polyakov loop, as in Eq.~\eqref{def:curvature}. Applying
the derivative operator 
$\partial_{\mu} \equiv \partial / \partial (\mu/T)$ to the numerator, which is
the only part depending on the chemical potential, one has
\begin{eqnarray}
  \partial_{\mu}^2\big|\langle\mathrm{Tr}L\rangle\big|^2
  &=& 2\left(\partial_{\mu}\langle\mathrm{ReTr}L\rangle\right)^2
      +2\langle\mathrm{ReTr}L\rangle\partial_{\mu}^2\langle\mathrm{ReTr}L\rangle
      \nonumber \\
  &+& \big\{\mathrm{ReTr}L\leftrightarrow\mathrm{ImTr}L\big\}~,
      \label{eq:d2l2}
  \end{eqnarray}
where $\mu=\mu_B/3$ is the common chemical potential for all
flavors, 
and the last line
in brackets indicates terms where real and imaginary parts of the
Polyakov loop are exchanged. The expectation values entering this
expression can be written 
as 
\begin{equation}
  \langle\mathrm{ReTr}L\rangle = \frac{1}{\mathcal{Z}}
  \int\hspace{-0.3em}\mathcal{D}U e^{- \mathcal{S}_{\text{YM}}}\mathrm{ReTr}L
  \prod_{f}\det\hspace{-0.2em}\left[ M_{\text{st}}^{f}\right]^{\frac{1}{4}},
  \label{eq:retrlaverage}
  \vspace{-0.3em}
\end{equation}
where a similar expression holds for
$\langle\mathrm{ImTr}L\rangle$ 
and $\mathcal{Z}$ is the partition function
defined in Eq.~\eqref{partfunc}.
Since the Polyakov loop does not depend explicitly on the
chemical potential, all dependence on $\mu$ is
carried by the Dirac matrix. That means that the derivative operator
will act only on the fermionic part of the
functional integral, which appears also in the denominator. One has
\begin{equation}
  \partial_{\mu}\prod_{f}\det\hspace{-0.2em}\left[ M_{\text{st}}^{f}\right]^{\frac{1}{4}} =
  \bigg(\sum_fn_f\bigg)\prod_{f}\det\hspace{-0.2em}\left[ M_{\text{st}}^{f}\right]^{\frac{1}{4}},
\end{equation}
where $n_f$ is the \emph{quark number} operators related to each
different flavor,
\begin{equation}
  n_f = \frac{1}{4}
  \mathrm{Tr}\left[{M_{\text{st}}^{f}}^{-1}
    \partial_{\mu} M_{\text{st}}^{f}\right] \, .
  \label{eq:numberdef}
\end{equation}
Setting $n = \sum_f n_f$ one can rewrite the derivative of the
expression in Eq.~\eqref{eq:retrlaverage} as
\begin{equation}
  \partial_{\mu}\langle\mathrm{ReTr}L\rangle
  = \langle n\, \mathrm{ReTr}L \rangle-
  \langle\mathrm{ReTr}L\rangle\langle n\rangle
  \label{eq:derretrl}
\end{equation}
and the same is true also for $\langle\mathrm{ImTr}L\rangle$. 
Further application of the derivative $\partial_{\mu}$ leads to 
new correlators 
involving the quark number $n$ or its derivative
$n'=\partial_{\mu}n$. Indeed, one finds that
\begin{eqnarray}
  \partial_{\mu}\langle n \, \mathrm{ReTr}L\rangle
  &=&
      \langle n^2 \, \mathrm{ReTr}L \rangle -
      \langle n \, \mathrm{ReTr}L\rangle\langle n\rangle + \langle
      n' \, \mathrm{ReTr}L \rangle \nonumber \\ 
  \partial_{\mu}\langle n\rangle
  &=&
      \langle n^2\rangle - \langle n\rangle^2 + \langle n'\rangle~,
\end{eqnarray}
where $n' = \sum_f n_f'$ and $n_f' = \partial_{\mu}n_f$ with
\begin{equation}
\partial_{\mu}n_f=
  \frac{1}{4}\mathrm{Tr}\left[\left({M_{\text{st}}^{f}}^{-1}
      \partial_{\mu} M_{\text{st}}^{f}\right)^2
    - {M_{\text{st}}^{f}}^{-1}
    \partial_{\mu}^2 M_{\text{st}}^{f}\right]~.
\end{equation}
Finally,
joining and re-arranging all the pieces appearing in
Eq.~\eqref{eq:d2l2}, the following expression is found
\begin{eqnarray}
  \partial_{\mu}^2|\langle\textnormal{Tr}L\rangle|^2
  &=&2\langle n \, \textnormal{ReTrL}\rangle^2+6\langle\textnormal{ReTrL}\rangle^2\langle n\rangle^2
      \nonumber\\
  &-&8\langle\textnormal{ReTrL}\rangle\langle n\, \textnormal{ReTrL}\rangle\langle n\rangle
      \nonumber\\
  &+&2\langle\textnormal{ReTrL}\rangle\langle n^2 \, \textnormal{ReTrL}\rangle
      -2\langle\textnormal{ReTrL}\rangle^2\langle n^2\rangle
      \nonumber\\
  &+&2\langle\textnormal{ReTrL}\rangle\langle n' \, \textnormal{ReTrL} \,\rangle
      -2\langle\textnormal{ReTrL}\rangle^2\langle n'\,\rangle
      \nonumber\\
  &+& \left\{\textnormal{ReTrL} \leftrightarrow
      \textnormal{ImTrL}\right\}~.
\end{eqnarray}
The curvature $\chi_{Q,\mu_B}$ is obtained by normalizing
this formula with the square module of $\langle\mathrm{Tr}L(0)\rangle$
and evaluating the ratio at zero chemical potential, see
Eq.~\eqref{def:curvature}. As a result, the expression above
simplifies since, for $\mu=0$, both the quark number
$\langle n\rangle$ and 
$\langle\mathrm{ImTr}L\rangle$ vanish because of charge
conjugation symmetry. Then, re-arranging the remaining terms the
definition in Eq.~\eqref{def:curvature2} is found.

%%%%%%%%%%%%%%%%%%%%%%%%%%%%%%%%%
%%%%%%%%% BIBLIOGRAPHY %%%%%%%%%%
%%%%%%%%%%%%%%%%%%%%%%%%%%%%%%%%%

\end{document}